\documentclass[12pt]{article}
\usepackage{amsmath}
\usepackage{amsfonts}
\usepackage{mitpress}

\newdimen\dummy
\dummy=\oddsidemargin
\begin{document}




\begin{center}
\bigskip {\Large Time-dependent coupled oscillator model for charged
particle motion in the presence of a time-varying magnetic field \vspace{%
0.7cm}}

$\mathbf{Salah\ Menouar}$\footnote{$^*$E-mail: menouar\_salah@yahoo.fr}$^*$,~%
$\mathbf{Mustapha\ Maamache}^{1}$,~$\mathbf{and\ Jeong\ Ryeol\ Choi}$%
\footnote{$^\dagger$Corresponding author, E-mail: choiardor@hanmail.net}$%
^\dagger$ \vspace{0.3cm}

\ $^{1}$\textit{Laboratoire de Physique Quantique et Syst\`{e}mes
Dynamiques, }

\textit{\ \ D\'{e}partement de Physique, Facult\'{e} des Sciences, }

\textit{\ \ Universit\'{e} Ferhat Abbas de S\'{e}tif, S\'{e}tif 19000,
Algeria \vspace{0.4cm}}


$\ \ ^{2}$\textit{Division of Semiconductor and Display Engineering, }

\ \ \textit{College of IT Engineering, Kyungpook National University, }

\ \ \ \textit{1370 Sankguk-dong, Buk-gu, Daegu 702-701, Republic of Korea
\vspace{0.4cm}}\ \ \

\end{center}

{\Large Abstract }

{\Large \bigskip }

The dynamics of time-dependent coupled oscillator model for the charged
particle motion subjected to a time-dependent external magnetic field is
investigated. We used canonical transformation approach for the classical
treatment of the system, whereas unitary transformation approach is used when
managing the system in the framework of quantum mechanics. For both
approaches, the original system is transformed to a much more simple system
that is the sum of two independent harmonic oscillators which have
time-dependent frequencies. We therefore easily identified the wave
functions in the transformed system with the help of invariant operator of
the system. The full wave functions in the original system is derived from
the inverse unitary transformation of the wave functions associated to the
transformed system. \newline
\newline
\textbf{PACC numbers}: 0365G, \ 0365D, 4190

\newpage

\section{\protect\bigskip Introduction}

The time-dependent harmonic oscillators have attracted considerable interest
in the literature thanks to their usefulness in describing the dynamics of
many physical systems. After the Bateman's\cite{bate} proposition concerning
the use of time-dependent harmonic oscillator model in describing
dissipative systems, much attention was paid to quantum behavior of
nonconservative and nonlinear systems.

In the meantime, coupled oscillators have emerged to become powerful
modeling tools and, consequently, are frequently used in modeling wide range of
physical phenomena. With the progress of research, one may be interested in
what would happen if two-dimensional harmonic oscillator is elaborated
through the coupling of two additive potentials? As far as we know,
dealing with such an issue was set thirty years ago by Kim et al. [2-5].
Abdalla demonstrated how to treat the time-dependent coupled oscillators in
the context of quantum mechanics\cite{mscb}. The propagator for a
time-dependent coupled and driven harmonic oscillators with time-varying
frequencies and masses is investigated by Benamira \cite{benami} using path
integral methods.

Among various systems that can be modeled by time-dependent coupled
oscillators, the dynamics of charged particle motion in the presence of
time-varying magnetic fields has played an important role in condensed
matter physics and plasma physics. There are plenty of applications for this
system such as magnetoresistance\cite{bykov}, the Aharonov-Bohm effect\cite%
{wgv}, magnetic confinement devices for fusion plasmas\cite{varma},
electromagnetic lenses with variable magnetic fields\cite{calvo}, cyclotron
resonance\cite{kenn}, and entanglement of a two-qubit Heisenberg XY model%
\cite{sadi}. Though all of these problems are interesting, we can find their
exact analytic solutions only for a few special cases due to their
complex mathematical structures.

The quantum properties {} of a free electron, {}which have a time-dependent
effective mass under the influence of external magnetic field, are
investigated in both the Landau and the symmetric gauges \cite{choi2,smen}.
Laroze and Rivera\cite{laroz} studied the dynamical behavior of electrons in
the presence of a uniform time-dependent magnetic field and they presented
the time evolution of the corresponding wave functions for the case that the initial
state is a superposition of Landau levels. The propagators of a charged
particle subjected to a time-dependent magnetic field are studied using the
linear and the quadratic invariants \cite{abdella3}.

Kim et al. [2-5] 
proposed a problem that what actually would take place if two harmonic
oscillators are coupled so that the potential becomes $V(X_{1},X_{2})=\frac{1%
}{2}\left( c_{1}X_{1}^{2}+c_{2}X_{2}^{2}+c_{3}X_{1}X_{2}\right) $ where $%
c_{3}$ is a coupling constant. They studied the corresponding density matrix
in order to establish the Wigner function. In this work, we are interested
in the problem of Hamiltonian that involves the coupling term $X_{1}X_{2}$
in the presence of magnetic field. This system can be regarded as the
generalization of the Hamiltonian model given in Refs. \cite{choi2} and \cite%
{gel1}. Though the coupling of two or more oscillators is among the most
basic concepts in dealing with gyroscopic motions, interactions, and complex
structures, the related theory  has been scarcely developed so far. This
class of coupled harmonic oscillators can be used to describe numerous
physical systems. Some of them are the Bogoliubov transformation model of
superconductivity \cite{kim7}, two-mode squeezed light \cite{cm}, and the
Lee model in quantum field theory \cite{ss}. One of the main focuses of
research carried out by Zhang \textit{et al}. in connection with
time-dependent coupled oscillators including $X_{1}X_{2}$ term are some
specific problems of time-dependent coupled electronic circuits\cite%
{choi5,choi6}.

We will use the invariant methods\cite{lewis3,lej} in order to derive the
exact wave functions for time-dependent coupled oscillators in a variable
magnetic field. The invariant operator method in describing the quantum features
of time-dependent harmonic oscillators is firstly introduced by Lewis\cite%
{lewis3} and now became a very useful tool in developing quantum theory
for the case where the Hamiltonian of the system is explicitly dependent on
time.

In Sec. 2, we formulate our problem by introducing a general time-dependent
Hamiltonian describing the complicated motion of a charged particle in the
presence of an arbitrary time-dependent magnetic field. Classical treatment
of the system is presented in Sec. 3, on the basis of the canonical
transformation method. Quantum analysis of the system is carried out in Sec.
4 using unitary transformation approach. The unitary transformation enables
us to transform the original Hamiltonian (that is somewhat complicated) to
that of a more simple system such as ordinary harmonic oscillator. We derive
the quantum solutions of the system in Sec. 5 starting from the invariant
operator associated to the transformed system described in Sec. 4. Finally,
we give concluding remarks in the last section.

\section{Formulation of the problem}

For the dynamical system of our interest, the Hamiltonian has the form:

\begin{equation}
H(X_{1},X_{2},t)=\frac{\Pi _{1}^{2}}{2m_{1}(t)}+\frac{\Pi _{2}^{2}}{2m_{2}(t)%
}+\frac{1}{2}\left(
C_{1}(t)X_{1}^{2}+C_{2}(t)X_{2}^{2}+C_{3}(t)X_{1}X_{2}\right) ,
\end{equation}%
where $\ \Pi _{1}$ and $\Pi _{2}$ are the conjugate momenta.
Note that $\ \Pi _{1}$ and $\Pi _{2}$ can be
simplified by choosing an appropriate gauge. Actually, in the symmetric
gauge with $\overrightarrow{A}\big(\frac{-B(t)}{2}X_{2},\frac{B(t)}{2}X_{1},0%
\big)$, they are given by%
\begin{equation}
\Pi _{1}=P_{1}-\frac{eB(t)}{2}X_{2}\ ,\ \Pi _{2}=P_{2}+\frac{eB(t)}{2}X_{1}.
\end{equation}%
The parameters $m_{1}(t),$ $m_{2}(t)$, $C_{1}(t)$, $C_{2}(t)$, and $C_{3}(t)$
are arbitrary functions of time, $(X_{1},X_{2})$ are the pair of position variables,
and $(P_{1}, P_{2})$ are the canonical conjugate momentum variables.

The main difference of our study from that of Ref. \cite{laroz} is that we
considered the coupling term $X_1 X_2$ in the Hamiltonian. Regarding the
expressions of $\ \Pi _{1}$ and\ \ $\Pi _{2}$, the Hamiltonian in Eq. (1)
can be recasted into

\begin{eqnarray}
H(X_{1},X_{2},t) &=&\frac{P_{1}^{2}}{2m_{1}(t)}+\frac{P_{2}^{2}}{2m_{2}(t)}+%
\frac{1}{2}\left(
c_{1}(t)X_{1}^{2}+c_{2}(t)X_{2}^{2}+c_{3}(t)X_{1}X_{2}\right)  \notag \\
&&+\frac{1}{2}\left( \omega _{2c}(t)P_{2}X_{1}-\omega
_{1c}(t)P_{1}X_{2}\right) ,
\end{eqnarray}%
where the new time-dependent functions $c_{1}(t)$, $c_{2}(t)$ and $c_{3}(t)$
are read
\begin{equation}
c_{1}(t)=C_{1}(t)+m_{2}(t)\frac{\omega _{2c}^{2}(t)}{4},\text{\ \ }%
c_{2}(t)=C_{2}(t)+m_{1}(t)\frac{\omega _{1c}^{2}(t)}{4},\text{\ \ }%
c_{3}(t)=C_{3}(t),
\end{equation}%
with the cyclotron frequencies
\begin{equation}
\omega _{1c}(t)=\frac{eB(t)}{m_{1}(t)},\text{ \ \ \ \ }\omega _{2c}(t)=\frac{%
eB(t)}{m_{2}(t)}.
\end{equation}

\section{Classical treatment}

The time-dependent canonical transformation approach is in fact very
powerful in investigating the properties of dynamical systems described by a
time-dependent Hamiltonian. In many cases, we can convert a given
Hamiltonian into a simple and desired one by means of the canonical
transformation. Therefore, in order to recast the solutions of this problem
into a more soluble form, it is convenient to use the canonical
transformation method. To simplify the Hamiltonian given in Eq. (3), let us
transform the variables $(X_{1},X_{2},P_{1},P_{2})$ \ to the new variables $%
(x_{1},x_{2},p_{x_{1}},p_{x_{2}})$ such that
\begin{equation}
x_{1}=\left( \frac{m_{1}(t)}{m_{2}(t)}\right) ^{1/4}X_{1},\text{ \ \ \ }%
x_{2}=\left( \frac{m_{2}(t)}{m_{1}(t)}\right) ^{1/4}X_{2},
\end{equation}%
\begin{equation}
p_{x_{1}}=\left( \frac{m_{2}(t)}{m_{1}(t)}\right) ^{1/4}P_{1},\text{ \ \ \ }%
p_{x_{2}}=\left( \frac{m_{1}(t)}{m_{2}(t)}\right) ^{1/4}P_{2}.
\end{equation}%
Replacing all of the canonical variables in Eq. (3) with the above ones, we have%
\begin{eqnarray}
H(x_{1},x_{2},t) &=&\frac{1}{2m(t)}\left( p_{x_{1}}^{2}+p_{x_{2}}^{2}\right)
+\frac{1}{2}\left(
d_{1}(t)x_{1}^{2}+d_{2}(t)x_{2}^{2}+d_{3}(t)x_{1}x_{2}\right)   \notag \\
&&+\frac{\omega _{c}(t)}{2}\left( x_{1}p_{x_{2}}-x_{2}p_{x_{1}}\right) ,
\end{eqnarray}%
where $d_{1}-d_{3}$ are new time-dependent functions of the form%
\begin{eqnarray}
d_{1}(t)&=&c_{1}(t)\left( \frac{m_{2}(t)}{m_{1}(t)}\right) ^{1/2}=\left(
C_{1}(t)+\frac{1}{4}m_{2}(t)\omega _{2c}^{2}(t)\right) \left( \frac{m_{2}(t)%
}{m_{1}(t)}\right) ^{1/2},~~~ \\
d_2(t)&=&c_2(t)\left( \frac{m_{1}(t)}{m_{2}(t)}\right) ^{1/2}=\left(
C_{2}(t)+\frac{1}{4}m_{1}(t)\omega _{1c}^{2}(t)\right) \left( \frac{m_{1}(t)%
}{m_{2}(t)}\right) ^{1/2},~~~
\end{eqnarray}%
\begin{equation}
d_{3}(t)=c_{3}(t)=C_{3}(t),
\end{equation}%
with the unique mass $m(t)=\left( m_{1}(t)m_{2}(t)\right) ^{1/2}$ and the
cyclotron frequency $\omega _{c}(t)=\left( \omega _{1c}(t)\omega
_{2c}(t)\right) ^{1/2}=eB(t)/m(t).$

To simplify the Hamiltonian of Eq. (8), we perform the following canonical
transformation
\begin{eqnarray}
\binom{x_{1}}{x_{2}}&=&\left(
\begin{array}{cc}
\cos \phi (t) & \sin \phi (t) \\
-\sin \phi (t) & \cos \phi (t)%
\end{array}%
\right) \binom{q_{1}}{q_{2}},  \\
\binom{p_{x_{1}}}{p_{x_{2}}}&=&\left(
\begin{array}{cc}
\cos \phi (t) & \sin \phi (t) \\
-\sin \phi (t) & \cos \phi (t)%
\end{array}%
\right) \binom{p_{1}}{p_{2}},
\end{eqnarray}%
where%
\begin{equation}
\phi (t)=-\frac{1}{2}\int \omega _{c}(t)dt.
\end{equation}%
If $(q_{1},q_{2},p_{1},p_{2})$ are canonical coordinates, there should exist
a new Hamiltonian $H(q_{1},q_{2},t)$ which is determined by only in terms of
the Hamiltonian given in Eq. (8) with the aid of the linear transformation
shown in Eqs. (12) and (13). The variables $\left(
x_{1},x_{2},p_{x_{1}},p_{x_{2}}\right) $ and $(q_{1},q_{2},p_{1},p_{2})$ in
two representations must satisfy the following relation \cite{gold}%
\begin{equation}
(p_{1}\dot{q}_{1}+p_{2}\dot{q}_{2}-H(q_{1},q_{2},t)=p_{x_{1}}\dot{x}%
_{1}+p_{x_{2}}\dot{x}_{2}-H(x_{1},x_{2},t)+\frac{\partial F_{1}}{\partial t},
\end{equation}%
where $F_{1}$ is a time-dependent generating function in phase space, which
should be determined afterwards.

From the fundamental equations known in classical mechanics \cite{gold}%
\begin{eqnarray}
p_{x_{1}} &=&\frac{\partial }{\partial x_{1}}F_{1}\left(
x_{1},x_{2},p_{1},p_{2},t\right) ,~~~~q_{1}=\frac{\partial }{\partial p_{1}}%
F_{1}\left( x_{1},x_{2},p_{1},p_{2},t\right) , \\
p_{x_{2}} &=&\frac{\partial }{\partial x_{2}}F_{1}\left(
x_{1},x_{2},p_{1},p_{2},t\right) ,~~~~q_{2}=\frac{\partial }{\partial p_{2}}%
F_{1}\left( x_{1},x_{2},p_{1},p_{2},t\right) ,
\end{eqnarray}%
the generating function associated with the transformation is found to be%
\begin{equation}
F_{1}\left( x_{1},x_{2},p_{1},p_{2}t\right) =\left( p_{1}\cos \phi
+p_{2}\sin \phi \right) x_{1}+\left( -p_{1}\sin \phi +p_{2}\cos \phi \right)
x_{2},
\end{equation}%
\begin{equation}
\frac{\partial F_{1}}{\partial t}=-\dot{\phi}(t)\left(
x_{1}p_{x_{2}}-x_{2}p_{x_{1}}\right) =-\frac{\varpi _{c}(t)}{2}\left(
x_{1}p_{x_{2}}-x_{2}p_{x_{1}}\right) .
\end{equation}%
In terms of the new conjugate variables $(q_{1},q_{2},p_{1},p_{2}),$ the
Hamiltonian of Eq. (8) becomes%
\begin{equation}
H(q_{1},q_{2},t)=\frac{1}{2m(t)}\left( p_{_{1}}^{2}+p_{_{2}}^{2}\right) +%
\frac{1}{2}\left( \lambda _{1}(t)q_{1}^{2}+\lambda _{2}(t)q_{2}^{2}+\lambda
_{3}(t)q_{1}q_{2}\right) ,
\end{equation}%
where%
\begin{eqnarray}
& &\lambda _{1}(t)=d_{1}(t)\cos ^{2}\phi +d_{2}(t)\sin ^{2}\phi -d_{3}(t)\sin
\phi \cos \phi \text{,}  \\
& &\lambda _{2}(t)=d_{2}(t)\cos ^{2}\phi +d_{1}(t)\sin ^{2}\phi +d_{3}(t)\sin
\phi \cos \phi \text{,}  \\
& &\lambda _{3}(t)=2\left( d_{1}(t)-d_{2}(t)\right) \sin \phi \cos \phi
+d_{3}(t)\left( \cos ^{2}\phi -\sin ^{2}\phi \right) \text{.}
\end{eqnarray}%
To eliminate the coupling term $q_{1}q_{2},$ we now perform the following
canonical transformation \cite{benami,choi5,choi6}
\begin{equation}
\binom{q_{1}}{q_{2}}=\frac{1}{\sqrt{m(t)}}\left(
\begin{array}{cc}
\cos \frac{\theta (t)}{2} & \sin \frac{\theta (t)}{2} \\
-\sin \frac{\theta (t)}{2} & \cos \frac{\theta (t)}{2}%
\end{array}%
\right) \binom{Q_{1}}{Q_{2}},
\end{equation}%
\begin{equation}
\binom{p_{1}}{p_{2}}=\sqrt{m(t)}\left(
\begin{array}{cc}
\cos \frac{\theta (t)}{2} & \sin \frac{\theta (t)}{2} \\
-\sin \frac{\theta (t)}{2} & \cos \frac{\theta (t)}{2}%
\end{array}%
\right) \binom{P_{1}}{P_{2}}-\left(
\begin{array}{cc}
\frac{\dot{m}(t)}{2} & 0 \\
0 & \frac{\dot{m}(t)}{2}%
\end{array}%
\right) \binom{q_{1}}{q_{2}}.
\end{equation}%
where $\theta(t)$ is an arbitrary function of time.
Note that Eqs. (24) and (25) do not always represent the canonical
transformation \cite{gold} between variables $\left( q_{i},p_{i}\right)
[i=1,2]$ and $\left( Q_{i},P_{i}\right) $. If $\left( Q_{i},P_{i}\right) $
are canonical coordinates, there should exist a new Hamiltonian which is
determined only by the Hamiltonian of Eq. (20) and the linear transformation
given in Eqs. (24) and (25). The relation between variables $\left(
q_{i},p_{i}\right) $ and $\left( Q_{i},P_{i}\right) $ in the two
representations are \cite{gold}
\begin{equation}
\sum_{i=1}^{2}P_{i}\dot{Q}_{i}-H_{Q}=\sum_{i=1}^{2}{}p_{i}\dot{q}_{i}-H_{q}+%
\frac{\partial F}{\partial t},
\end{equation}%
where $F$ is an another time-dependent generating function in phase space.

Using the basic equations
\begin{equation}
p_{i}=\frac{\partial }{\partial q_{i}}F\left(
q_{1},q_{2},P_{1},P_{2},t\right) ,~~~Q_{i}=\frac{\partial }{\partial P_{i}}%
F\left( q_{1},q_{2},P_{1},P_{2},t\right) ,
\end{equation}%
where $i=1,2,$ we see that the generating function is given by%
\begin{eqnarray}
F\left( q_{1},q_{2},P_{1},P_{2},t\right) &=&\sqrt{m(t)}\left( P_{1}\cos
\frac{\theta (t)}{2}+P_{2}\sin \frac{\theta (t)}{2}\right) q_{1}  \notag \\
&&+\sqrt{m(t)}\left( -P_{1}\sin \frac{\theta (t)}{2}+P_{2}\cos \frac{\theta
(t)}{2}\right) q_{2}  \notag \\
&&-\frac{1}{4}\dot{m}(t)\left( q_{1}^{2}+q_{2}^{2}\right) .
\end{eqnarray}%
Then, in terms of the new conjugate variables $\left( Q_{i},P_{i}\right) $,
the Hamiltonian can be represented in the form
\begin{eqnarray}
H_{Q}(Q_{1},Q_{2},t) &=&\frac{1}{2}\left( P_{1}^{2}+P_{2}^{2}\right) +\frac{1%
}{2}\Omega _{1}^{2}(t)Q_{1}^{2}+\frac{1}{2}\Omega _{2}^{2}(t)Q_{2}^{2}
\notag \\
&&+\frac{\dot{\theta}(t)}{2}\left[ P_{1}Q_{2}-P_{2}Q_{1}\right] +\delta
(t)Q_{1}Q_{2}.
\end{eqnarray}%
Here, the time-dependent coefficients $\Omega _{1}(t),\Omega _{2}(t)$ and $%
\delta (t)$ are given by%
\begin{equation}
\Omega _{1}(t)=\left( \tilde{\omega}_{1}^{2}(t)\cos ^{2}\frac{\theta (t)}{2}+%
\tilde{\omega}_{2}^{2}(t)\sin ^{2}\frac{\theta (t)}{2}-\frac{\lambda
_{3}(t)\sin \theta (t)}{m(t)}\right) ^{1/2},
\end{equation}%
\begin{equation}
\Omega _{2}(t)=\left( \tilde{\omega}_{1}^{2}(t)\sin ^{2}\frac{\theta (t)}{2}+%
\tilde{\omega}_{2}^{2}(t)\cos ^{2}\frac{\theta (t)}{2}+\frac{\lambda
_{3}(t)\sin \theta (t)}{m(t)}\right) ^{1/2},
\end{equation}%
\begin{equation}
\delta (t)=\frac{1}{2}\left( \tilde{\omega}_{1}^{2}(t)-\tilde{\omega}%
_{2}^{2}(t)\right) \sin \theta (t)+\frac{\lambda _{3}(t)\cos \theta (t)}{m(t)%
},
\end{equation}%
where%
\begin{eqnarray}
\tilde{\omega}_{1}^{2}(t) &=&\frac{\lambda _{1}(t)}{m(t)}+\frac{1}{4}\left(
\frac{\dot{m}^{2}(t)}{m^{2}(t)}-2\frac{\ddot{m}(t)}{m(t)}\right) , \\
\tilde{\omega}_{2}^{2}(t) &=&\frac{\lambda _{2}(t)}{m(t)}+\frac{1}{4}\left(
\frac{\dot{m}^{2}(t)}{m^{2}(t)}-2\frac{\ddot{m}(t)}{m(t)}\right) .
\end{eqnarray}

If we take the choice $\theta (t)=\mathrm{Const},$ the terms $P_{1}Q_{2}$
and $P_{2}Q_{1}$ in Eq. (29) are canceled out so that the Hamiltonian becomes%
\begin{equation}
H_{Q}(Q_{1},Q_{2},t)=\frac{1}{2}\left( P_{1}^{2}+P_{2}^{2}\right) +\frac{1}{2%
}\Omega _{1}^{2}(t)Q_{1}^{2}+\frac{1}{2}\Omega _{2}^{2}(t)Q_{2}^{2}+\delta
(t)Q_{1}Q_{2}.
\end{equation}

Notice that, with the above canonical transformation, the coupling $\delta
(t)$ is a functional on the parameters of the original system. It is hence
clear that the separation of variables in Eq. (35) requires that $\delta
(t)=0$, i.e.%
\begin{equation}
\lambda _{3}(t)=\left( \tilde{\omega}_{2}^{2}(t)-\tilde{\omega}%
_{1}^{2}(t)\right) m(t)\tan \theta ,
\end{equation}%
and consequently%
\begin{equation}
\tan \theta =\frac{\lambda _{3}(t)}{m(t)\left( \tilde{\omega}_{2}^{2}(t)-%
\tilde{\omega}_{1}^{2}(t)\right) }.
\end{equation}

By taking into account Eq. (36), the Hamiltonian in Eq. (35) is rewritten as%
\begin{equation}
H_{Q}(Q_{1},Q_{2},t)=\frac{1}{2}\left( P_{1}^{2}+P_{2}^{2}\right) +\frac{1}{2%
}\Omega _{1}^{2}(t)Q_{1}^{2}+\frac{1}{2}\Omega _{2}^{2}(t)Q_{2}^{2}.
\end{equation}%
Then, Eq.(38) represents the sum of two independent Hamiltonians of the
simple harmonic oscillators with the time-dependent frequencies $\ \Omega
_{1}(t)$ and $\Omega _{2}(t)$.

\section{Quantum treatment}

The canonical transformations in classical mechanics, treated in the
previous section, is the analogous of the unitary transformations in quantum
mechanics. Now we are going to demonstrate this relationship between the two
transformations and confirm how to obtain the quantum-mechanical Hamiltonian
from the classical one. To manage the system in the context of quantum
physics, we replace the canonical variables $\left( X_{1},X_{2}\right) $  in Eq. (3) by
quantum operators $(\hat{X}_{1},\hat{X}_{2})$. Then the corresponding
Hamiltonian has the form

\begin{eqnarray}
\hat{H}(\hat{X}_{1},\hat{X}_{2},t) &=&\frac{\hat{P}_{1}^{2}}{2m_{1}(t)}+%
\frac{\hat{P}_{2}^{2}}{2m_{2}(t)}+\frac{1}{2}\left( c_{1}(t)\hat{X}%
_{1}^{2}+c_{2}(t)\hat{X}_{2}^{2}+c_{3}(t)\hat{X}_{1}\hat{X}_{2}\right)
\notag \\
&&+\frac{1}{2}\left( \omega _{2c}(t)\hat{P}_{2}\hat{X}_{1}-\omega _{1c}(t)%
\hat{P}_{1}\hat{X}_{2}\right) .
\end{eqnarray}%
In this quantum case, the pair of momentum operators are given by $(\hat{P}_{1}=-i\hbar
\partial /\partial X_{1}$, $\hat{P}_{2}=-i\hbar \partial /\partial X_{2})$.
The Schr\"{o}dinger equation in the original system is

\begin{equation}
i\hbar \frac{\partial }{\partial t}\Psi (X_{1},X_{2},t)=\hat{H}(\hat{X}_{1},%
\hat{X}_{2},t)\Psi (X_{1},X_{2},t).
\end{equation}

To simplify the Hamiltonian in Eq. (39), we perform the unitary
transformation such that

\begin{equation}
\Psi (X_{1},X_{2},t)=\hat{U}_{1}(t)\psi (X_{1},X_{2},t),
\end{equation}%
where $\hat{U}_{1}(t)$ is a time-dependent unitary operator of the form%
\begin{eqnarray}
\hat{U}_{1}(t) &=&\exp \frac{i}{2\hbar }\left[ (\hat{P}_{1}\hat{X}_{1}+\hat{X%
}_{1}\hat{P}_{1})\ln \left( \frac{m_{1}(t)}{m_{2}(t)}\right) ^{1/4}\right]
\notag \\
&&\times\exp \frac{i}{2\hbar }\left[ (\hat{P}_{2}\hat{X}_{2}+\hat{X}_{2}\hat{%
P}_{2})\ln \left( \frac{m_{2}(t)}{m_{1}(t)}\right) ^{1/4}\right].
\end{eqnarray}%
In this case, the Hamiltonian, Eq. (39), can be rewritten as%
\begin{eqnarray}
& &\hat{H}_{1}(\hat{X}_{1},\hat{X}_{2},t) =\frac{1}{2m(t)}\left( \hat{P}%
_{1}^{2}+\hat{P}_{2}^{2}\right)  \notag \\
& &~~~~~~~~~~~+\frac{1}{2}\left( d_{1}(t)\hat{X}_{1}^{2}+d_{2}(t)\hat{X}%
_{2}^{2}+d_{3}(t)\hat{X}_{1}\hat{X}_{2}\right)  \notag \\
& &~~~~~~~~~~~+\frac{\omega _{c}(t)}{2}\left( \hat{P}_{2}\hat{X}_{1}-\hat{P}%
_{1}\hat{X}_{2}\right).
\end{eqnarray}

It is easy to confirm that the commutation relations, $[\hat{L}_{Z}\text{ },%
\text{ }\hat{X}_{1}^{2}+\hat{X}_{2}^{2}]=0$ and $[\hat{L}_{z}\text{ },\text{
}\hat{P}_{1}^{2}+\hat{P}_{2}^{2}]=0$, are hold where $\hat{L}_{Z}$ is the
angular momentum operator. This implies that there are common eigenfunctions
between $\hat{L}_{Z}$ and $\hat{X}_{1}^{2}+\hat{X}_{2}^{2}$, and between $%
\hat{L}_{Z}$ and $\hat{P}_{1}^{2}+\hat{P}_{2}^{2}$. However, $\hat{L}_{Z}$
does not commutes with $\hat{X}_{1}\hat{X}_{2}$: $[\hat{L}_{Z}\text{ },\text{
}\hat{X}_{1}\hat{X}_{2}]\neq 0$, and consequently $[\hat{L}_{Z}\text{ },%
\text{ }\hat{H}]\neq 0$. If we regard that $\hat{L}_{Z}$ and $\hat{H}$ do
not have the same eigenfunctions, it is not possible to simplify the Schr%
\"{o}dinger equation
\begin{equation}
i\hbar \frac{\partial }{\partial t}\psi (X_{1},X_{2},t)=\hat{H}_{1}(\hat{X}%
_{1},\hat{X}_{2},t)\psi (X_{1},X_{2},t),
\end{equation}%
by decomposing it. However, we can overcome this difficulty through the
transformation of the Hamiltonian of Eq. (39) into a simple form by
introducing an appropriate unitary transformation operators. In the first
step, we perform the following unitary transformation%
\begin{equation}
\psi (X_{1},X_{2},t)=\hat{U}_{2}(t)\varphi (X_{1},X_{2},t),
\end{equation}%
where%
\begin{eqnarray}
\hat{U}_{2}(t) &=&\exp \left( -\frac{i}{2\hbar }\left( \hat{P}_{2}\hat{X}%
_{1}-\hat{P}_{1}\hat{X}_{2}\right) \int \varpi _{c}(t)dt\right)   \notag \\
&=&\exp \left( -\frac{i\hat{L}_{Z}}{2\hbar }\int \varpi _{c}(t)dt\right) .
\end{eqnarray}%
Under this transformation, the Schr\"{o}dinger equation (41) is mapped into%
\begin{equation}
i\hbar \frac{\partial }{\partial t}\varphi (X_{1},X_{2},t)=\hat{H}_{2}(\hat{X%
}_{1},\hat{X}_{2},t)\varphi (X_{1},X_{2},t),
\end{equation}%
where the new Hamiltonian $\hat{H}_{2}(\hat{X}_{1},\hat{X}_{2},t)$ has the
form%
\begin{eqnarray}
&&\hat{H}_{2}(\hat{X}_{1},\hat{X}_{2},t)=\frac{1}{2m(t)}\left( \hat{P}%
_{1}^{2}+\hat{P}_{2}^{2}\right)   \notag \\
&&~~~~~~~~~~+\frac{1}{2}\left( \lambda _{1}(t)\hat{X}_{1}^{2}+\lambda _{2}(t)%
\hat{X}_{2}^{2}+\lambda _{3}(t)\hat{X}_{1}\hat{X}_{2}\right) .
\end{eqnarray}%
Now the term involving $\hat{L}_{Z}$ has disappeared in Eq. (48). This means
that the magnetic field term is removed in the new frame rotating with the
time-dependent phase $\phi (t)=-\frac{1}{2}\int \varpi _{c}(t)dt.$

To decouple the Hamiltonian of Eq. (48), we take another unitary
transformation such that
\begin{equation}
\varphi (X_{1},X_{2},t)=\hat{V}(t)\chi (X_{1},X_{2},t),
\end{equation}%
where the unitary operator $\hat{V}(t)$ is given by%
\begin{equation}
\hat{V}(t)=\hat{V}_{1}(t)\hat{V}_{2}(t)\hat{V}_{3}(t),
\end{equation}%
with
\begin{eqnarray}
\hat{V}_{1}(t) &=&\exp \frac{i}{2\hbar }\left[ (\hat{P}_{1}\hat{X}_{1}+\hat{X%
}_{1}\hat{P}_{1})\ln \sqrt{m(t)}\right]  \notag \\
&&\times \exp \frac{i}{2\hbar }\left[ (\hat{P}_{2}\hat{X}_{2}+\hat{X}_{2}%
\hat{P}_{2})\ln \sqrt{m(t)}\right] , \\
\hat{V}_{2}(t) &=&\exp \left[ -\frac{i}{\hbar }\frac{\theta }{2}(\hat{P}_{2}%
\hat{X}_{1}-\hat{P}_{1}\hat{X}_{2})\right] , \\
\hat{V}_{3}(t) &=&\exp -\frac{i}{4\hbar }\dot{m}(t)\left( \hat{X}_{1}^{2}+%
\hat{X}_{2}^{2}\right) .
\end{eqnarray}%
Some algebra with the substitution of Eqs. (48) and (49) into Eq.(47) yields
a transformed Hamiltonian that represents the sum of two uncoupled simple
harmonic oscillators having frequencies $\Omega _{1}(t)$ and $\Omega _{2}(t)$
and the unit mass:
\begin{eqnarray}
\hat{H}_{3}(\hat{X}_{1},\hat{X}_{2},t) &=&\hat{V}^{-1}(t)\hat{H}_{2}(\hat{X}%
_{1},\hat{X}_{2},t)\hat{V}(t)-i\hbar \hat{V}^{-1}(t)\frac{\partial }{%
\partial t}\hat{V}(t)  \notag \\
&=&\frac{1}{2}\left( \hat{P}_{1}^{2}+\hat{P}_{2}^{2}\right) +\frac{1}{2}%
\Omega _{1}^{2}(t)\hat{X}_{1}^{2}+\frac{1}{2}\Omega _{2}^{2}(t)\hat{X}%
_{2}^{2}.
\end{eqnarray}%
At this stage, it is possible to confirm that the classically transformed
Hamiltonian given in Eq. (38) is right, since the above equation is
consistent with it. Note that $\hat{U}_{1}(t)$ and$\ \hat{V}_{1}(t)$ given
in Eqs. (42) and (51) are the squeeze operators, whereas $\hat{U}_{2}(t)$ and%
$\ \hat{V}_{2}(t)$ given in Eqs. (46) and (52) are the rotation operators
characterized by the time-varying angles $\phi (t)$ and $\frac{\theta (t)}{2}
$, respectively.

\section{Quantum solutions}

It can be seen that there exists invariant for the harmonic oscillator with
time-dependent mass and/or frequency\cite{lewis3}. In our case, the
transformed system consists of the two independent harmonic oscillators
which have time-dependent frequency. It is easy to verify, from Liouville-von
Neumann equation for the invariant $\hat{I}$
\begin{equation}
\frac{d\hat{I}}{dt}=\frac{\partial \hat{I}}{\partial t}+\frac{1}{i\hbar }[%
\hat{I},\hat{H}_{3}]=0,
\end{equation}%
that the invariant associated to the Hamiltonian of two-dimensional harmonic
oscillator is given by%
\begin{eqnarray}
\hat{I}(\hat{X}_{1},\hat{X}_{2},t) &=&\hat{I}(\hat{X}_{1},t)+\hat{I}(\hat{X}%
_{2},t)  \notag \\
&=&\frac{1}{2}\left[ \left( \frac{\hat{X}_{1}}{\rho _{1}}\right) ^{2}+\left( \rho
_{1}\overset{\cdot }{\hat{X}_{1}}-\dot{\rho}_{1}\hat{X}_{1}\right) ^{2}\right]
\notag \\
&&+\frac{1}{2}\left[ \left( \frac{\hat{X}_{2}}{\rho _{2}}\right) ^{2}+\left( \rho
_{2}\overset{\cdot }{\hat{X}_{2}}-\dot{\rho}_{2}\hat{X}_{2}\right) ^{2}\right] ,
\end{eqnarray}%
where $\rho _{1}(t)$ and $\rho _{2}(t)$ are c-number quantities obeying the
auxiliary equations

\begin{eqnarray}
\ddot{\rho}_{1}+\Omega _{1}^{2}(t)\rho _{1} &=&1/\rho _{1}^{3}, \\
\ddot{\rho}_{2}+\Omega _{2}^{2}(t)\rho _{2} &=&1/\rho _{2}^{3}.
\end{eqnarray}%
To guarantee
the Hermiticity of Eq. (56) ($\hat{I}^{\dagger }=\hat{I}$), we choose only the real
solutions of the above two equations.
It is clear that $\hat{I}(\hat{X}_{1},\hat{X}_{2},t)$ satisfies the
Liouville-Von Neumann equation.
We now derive a complete orthonormal set of eigenfunctions $\xi
_{n_{1}n_{2}}(X_{1},X_{2},t)$ of $\hat{I}(\hat{X}_{1},\hat{X}_{2},t)$ form
the eigenvalue equation%
\begin{equation}
\hat{I}(\hat{X}_{1},\hat{X}_{2},t)\xi _{n_{1}n_{2}}(X_{1},X_{2},t)=\lambda
_{n_{1}n_{2}}\xi _{n_{1}n_{2}}(X_{1},X_{2},t),
\end{equation}%
where $\lambda _{n_{1}n_{2}}$ are time-{\it in}dependent eigenvalues.
Through a straightforward evaluation after inserting Eq. (56) into the above equation,
we get the eigenvalues and the eigenfunctions such that%
\begin{equation}
\lambda _{n_{1}n_{2}}=\hbar \left( n_{1}+\frac{1}{2}\right) +\hbar \left(
n_{2}+\frac{1}{2}\right) ,
\end{equation}%
\begin{eqnarray}
&&\xi _{n_{1}n_{2}}(X_{1},X_{2},t)=\left[ \frac{1}{\pi \hbar
n_{1}!n_{2}!2^{n_{1}+n_{2}}\rho _{1}\rho _{2}}\right] ^{1/2}  \notag \\
&&~~~~~~~~~\times H_{n_{1}}\left( \frac{X_{1}}{\hbar ^{1/2}\rho _{1}}\right)
H_{n_{2}}\left( \frac{X_{2}}{\hbar ^{1/2}\rho _{2}}\right)   \notag \\
&&~~~~~~~~~\times \exp \left[ \frac{i}{2\hbar }\left( \frac{\dot{\rho}_{1}}{%
\rho _{1}}+\frac{i}{\rho _{1}^{2}}\right) X_{1}^{2}+\frac{i}{2\hbar }\left(
\frac{\dot{\rho}_{2}}{\rho _{2}}+\frac{i}{\rho _{2}^{2}}\right) X_{2}^{2}%
\right] ,
\end{eqnarray}%
where $H_{n_{1}}$ and $H_{n_{2}}$ are the usual Hermite polynomial of order $%
n_{1}$ and $n_{2}$ respectively.

The solutions of the Schr\"{o}dinger equation%
\begin{equation}
i\hbar \frac{\partial \chi _{n_{1}n_{2}}(X_{1},X_{2},t)}{\partial t}=\hat{H}%
_{3}(\hat{X}_{1},\hat{X}_{2},t)\chi _{n_{1}n_{2}}(X_{1},X_{2},t),
\end{equation}%
can be written as%
\begin{equation}
\chi _{n_{1}n_{2}}(X_{1},X_{2},t)=e^{i\alpha _{n_{1}n_{2}}(t)}\xi
_{n_{1}n_{2}}(X_{1},X_{2},t),
\end{equation}%
where the phase functions $\alpha _{n_{1}n_{2}}(t)$ satisfy the equation%
\begin{equation}
\frac{\partial }{\partial t}\alpha _{n_{1}n_{2}}(t)=\frac{1}{\hbar }%
\left\langle \xi _{n_{1}n_{2}}(X_{1},X_{2},t)\right\vert \frac{\partial }{%
\partial t}-\hat{H}_{3}(\hat{X}_{1},\hat{X}_{2},t)\left\vert \xi
_{n_{1}n_{2}}(X_{1},X_{2},t)\right\rangle .
\end{equation}

According to Eqs. (61) and (63), the solutions $\chi_{n_{1}n_{2}} (X_{1},X_{2},t)$
of the Schr\"{o}dinger equation (62), in the transformed system, becomes
\begin{eqnarray}
& &\chi _{n_{1}n_{2}}(X_{1},X_{2},t) = e^{i\alpha _{n_{1}n_{2}}(t)}\left[
\frac{1}{\pi \hbar n_{1}!n_{2}!2^{n_{1}+n_{2}}\rho _{1}\rho _{2}}\right]
^{1/2}  \notag \\
&&~~~~~~~~~~~\times H_{n_{1}}\left( \frac{X_{1}}{\hbar ^{1/2}\rho _{1}}%
\right) H_{n_{2}}\left( \frac{X}{\hbar ^{1/2}\rho _{2}}\right)  \notag \\
&&~~~~~~~~~~~\times \exp \left[ \frac{i}{2\hbar }\left( \frac{\dot{\rho}_{1}%
}{\rho _{1}}+\frac{i}{\rho _{1}^{2}}\right) X_{1}^{2}+\frac{i}{2\hbar }%
\left( \frac{\dot{\rho}_{2}}{\rho _{2}}+\frac{i}{\rho _{2}^{2}}\right)
X_{2}^{2}\right] ,
\end{eqnarray}%
where the time-dependent phase functions are given by
\begin{equation}
\alpha _{n_{1}n_{2}}(t)=-\left( n_{1}+\frac{1}{2}\right) \int_{0}^{t}\frac{%
dt^{\prime }}{\rho _{1}^{2}(t^{\prime })}-\left( n_{2}+\frac{1}{2}\right)
\int_{0}^{t}\frac{dt^{\prime }}{\rho _{2}^{2}(t^{\prime })}.
\end{equation}

The relation between the wave functions, $\Psi _{n_{1}n_{2}}(X_{1},X_{2},t)$%
, in the original system described by the Hamiltonian of Eq. (3)\ and the
wave functions $\chi _{n_{1}n_{2}}(X_{1},X_{2},t)$\ in the transformed
system is%
\begin{eqnarray}
\Psi _{n_{1}n_{2}}(X_{1},X_{2},t) &=&\hat{U}_{1}(t)\hat{U}_{2}(t)\hat{V}%
(t)\chi _{n_{1}n_{2}}(X_{1},X_{2},t)  \notag \\
&=&\hat{U}_{1}(t)\hat{U}_{2}(t)\hat{V}_{1}(t)\hat{V}_{2}(t)\hat{V}%
_{3}(t)\chi _{n_{1}n_{2}}(X_{1},X_{2},t).
\end{eqnarray}%
Using Eqs. (42), (46), (50) and (65), we derive the full wave functions
in the form%
\begin{eqnarray}
&&\Psi _{n_{1}n_{2}}(X_{1},X_{2},t)=\left[ \frac{\sqrt{m_{1}m_{2}}}{\pi
\hbar n_{1}!n_{2}!2^{n_{1}+n_{2}}\rho _{1}\rho _{2}}\right] ^{1/2}  \notag \\
&&~~~~~~\times H_{n_{1}}\left( \frac{\sqrt{m_{1}}\cos \left( \phi +\theta
/2\right) X_{1}-\sqrt{m_{2}}\sin \left( \phi +\theta /2\right) X_{2}}{\hbar
^{1/2}\rho _{1}}\right)   \notag \\
&&~~~~~~\times H_{n_{2}}\left( \frac{\sqrt{m_{1}}\sin \left( \phi +\theta
/2\right) X_{1}+\sqrt{m_{2}}\cos \left( \phi +\theta /2\right) X_{2}}{\hbar
^{1/2}\rho _{2}}\right)   \notag \\
&&~~~~~~\times \exp \frac{im_{1}}{2\hbar }\left( \frac{\gamma }{2}+\frac{%
\beta }{2}+\left( \frac{\beta }{2}-\frac{\gamma }{2}\right) \sin \left(
\theta +2\phi \right) \right) X_{1}^{2}  \notag \\
&&~~~~~~\times \exp \frac{im_{2}}{2\hbar }\left( \frac{\gamma }{2}+\frac{%
\beta }{2}-\left( \frac{\beta }{2}-\frac{\gamma }{2}\right) \sin \left(
\theta +2\phi \right) \right) X_{2}^{2}  \notag \\
&&~~~~~~\times \exp \frac{i}{2\hbar }\sqrt{m_{1}m_{2}}\left( \left( \beta
-\gamma \right) \cos \left( \theta +2\phi \right) \right) X_{1}X_{2}  \notag
\\
&&~~~~~~\times \exp i\left[ -\left( n_{1}+\frac{1}{2}\right) \int_{0}^{t}%
\frac{dt^{\prime }}{\rho _{1}^{2}(t^{\prime })}-\left( n_{2}+\frac{1}{2}%
\right) \int_{0}^{t}\frac{dt^{\prime }}{\rho _{2}^{2}(t^{\prime })}\right] ,
\end{eqnarray}%
where the time-dependent coefficients $\gamma (t)$ and $\beta (t)$ are given
as%
\begin{equation}
\gamma (t)=\left( \frac{\dot{\rho}_{1}}{\rho _{1}}+\frac{i}{\rho _{1}^{2}}-%
\frac{1}{2}\frac{d}{dt}\sqrt{m_{1}m_{2}}\right) ,
\end{equation}%
\begin{equation}
\beta (t)=\left( \frac{\dot{\rho}_{2}}{\rho _{2}}+\frac{i}{\rho _{2}^{2}}-%
\frac{1}{2}\frac{d}{dt}\sqrt{m_{1}m_{2}}\right) .
\end{equation}%
The full solutions in the original system, given in Eq. (68), are exact
since we did not use approximation or perturbation methods. Though these
solutions are somewhat complicated, they are very useful in predicting the
quantum behavior of the system. A merit of such analytical solutions is that
they can be employed in deriving the evolution of the probability
distribution, regardless of the change of the system's parameters. However,
the numerical solutions in this field, such as the one obtained from FDTD
(finite difference time domain) method\cite{fdtd}, are somewhat inconvenient
as inputs to further analyses, since one should recalculate the results
whenever the parameters of the system changes. Using Eq. (68), one can
easily take a complete description of the charged particle motion even when
the parameters of the system vary from time to time provided that the
classical solutions of Eqs. (57) and (58) are known.

\section{Conclusion}

We investigated the quantal problem of the time-dependent coupled oscillator
model associated to the charged particle motion in the presence of
time-dependent magnetic field. Though the behavior of charged particle in
magnetic field drew great concern in both quantum and classical view point,
researches in this line are rather concentrated on static problems that can
be modeled by time-\textit{in}dependent harmonic oscillator.

The system we treated in this work is however a more generalized one. It is
summarized as follows: \newline
(i) We supposed that the effective mass of the charged particle varies
explicitly with time under the influence of the time-dependent magnetic
field. If electrons or holes in the condensed matter interact with
environment or various excitations such as pressure, energy, temperature,
and stress, their effective mass may naturally vary with time\cite{choi2}.
Moreover, the random changes of the external field in the heterojunctions
and solid solutions give rise to the variation of effective mass in
accordance with the fluctuation of the composition in the system\cite{zsg}.
\newline
(ii) We let the external magnetic field $B(t)$ be an \textit{arbitrary}
function of time. Therefore, the application of our theory is not confined
in a special system that has a specific class of time-dependence for $B(t)$.
In fact, we can apply it in wide range of practical systems with the
flexible choice of the type of $B(t)$. \newline
(iii) Our system is further generalized by adding a coupling term $X_{1}X_{2}$ in
the Hamiltonian.

Through these generalization, the system became a somewhat complicated one
that is described in terms of time-dependent Hamiltonian. Since the
treatment of the original Hamiltonian system is not an easy task in this
case, we transformed our system to that of a much more simplified one using
two different techniques. In the first one, we carried out canonical
transformations in order to simplify the problem relevant to the original
classical Hamiltonian given in Eq. (1). After the transformation, the
Hamiltonian reduced to a simple form associated to two uncoupled harmonic
oscillators that each have time-dependent frequencies $\Omega _{1}(t)$ and $%
\Omega _{2}(t)$. In the second technique we used an alternative approach on
the basis of the unitary transformation method. With the choice of unitary
operators $\hat{U}_{1}(t)$, $\hat{U}_{2}(t)$ and $\hat{V}(t)$, the quantum
Hamiltonian (39) has been transformed to an equally simple one as that of
the canonical transformation previously performed, but within the realm of
quantum mechanics.

Since the Hamiltonian in the transformed system is very simple, we easily
constructed dynamical invariant operator $\hat{I}(\hat{X}_1,\hat{X}_2,t)$
associated to the transformed system, as given in Eq. (55). The eigenstates $%
\xi _{n_{1}n_{2}}(X_{1},X_{2},t)$ of this invariant operator are represented
in terms of the Hermite polynomial. The Schr\"{o}dinger solutions $\chi
_{n_{1}n_{2}}(X_{1},X_{2},t)$ in the transformed system are the same as $\xi
_{n_{1}n_{2}}(X_{1},X_{2},t)$ except for the time-dependent phase factor $%
e^{i\alpha _{n_{1}n_{2}}(t)}$. From the inverse transformation of $%
\chi_{n_{1}n_{2}} (X_{1},X_{2},t)$ with the unitary operators, we derived
the full wave functions (quantum solutions) in the original system [see Eq.
(68)]. The quantum solutions are expressed in terms of $\rho_1$ and $\rho_2$
that are the two independent solutions of the classical equation of motion
given in Eqs. (56) and (57), respectively. Even if we represented the
quantum solutions in terms of the classical solutions associated with the
\textit{transformed system}, it is also possible to represent them in terms
of the classical solutions associated with \textit{original system}. The
wave functions given in Eq. (68) can be used to investigate various quantum
properties of the system such as the fluctuations of canonical variables,
the evolution of quantum energy, and probability densities, even when the
parameters of the system vary from time to time. This is the advantage of
such analytical solutions over numerical solutions obtained, for example,
using the FDTD method\cite{fdtd}. \newline

\textbf{Acknowledgements} \newline
The work of J. R. Choi was supported by Basic Science Research Program through the National Research Foundation of
Korea(NRF) funded by the Ministry of Education, Science and Technology (No. 2010-0016914). \newline

\end{document}